\documentclass[prb,showpacs,twocolumn,preprintnumbers,amsmath,amssymb,floatf]{revtex4-1}

\usepackage{lipsum}
\usepackage[dvips]{graphicx}
\usepackage[latin1]{inputenc}
\usepackage{stmaryrd}
\usepackage{wasysym}
\usepackage{subfigure}
\usepackage{booktabs}

\usepackage{color}
\usepackage{epstopdf}
\usepackage{pdfpages}

\begin{document}
\draft
\title{Non-conventional superconducting fluctuations in Ba(Fe$_{1-x}$Rh$_x$)$_2$As$_2$ iron-based superconductors}
\author{L. Bossoni,$^{1}$  L. Roman\'o,$^{2}$ P.C.
Canfield,$^{3}$ A. Lascialfari$^{4}$}

\affiliation{$^{1}$ Department of Physics, University of
Pavia-CNISM, Via Bassi 6, I-27100 Pavia, Italy}
\address{$^{2}$ Department of Physics, University of Parma, Parco Area delle Scienze 7A, I-43100 Parma, Italy}
\address{$^{3}$ Ames Laboratory US DOE and Department of Physics and
Astronomy, Iowa State University, Ames, IA 50011, USA}
\address{$^{4}$ Dipartimento di Fisica and INSTM, Universit\'a degli Studi di Milano, Via Celoria 16, I-20133 Milano, Italy}

\begin{abstract}

We measured the static uniform spin susceptibility of Ba(Fe$_{1-x}$Rh$_x$)$_2$As$_2$ iron-based superconductors, over a broad range of doping ($0.041\leq x\leq 0.094$) and magnetic fields. At small fields ($H \le$ 1 kOe) we observed, above the transition temperature $T_c$, the occurrence of precursor diamagnetism, which is not ascribable to the Ginzburg-Landau theory. On the contrary, our data fit a phase fluctuation model, which has been used to interpret a similar phenomenology occurring in the high-$T_c$ cuprate superconductors. On the other hand, in presence of strong fields the unconventional fluctuating diamagnetism is suppressed, whereas 3D fluctuations are found, in agreement with literature.
\end{abstract}

 \maketitle

%%%%%%%%%%%%%%INTRO%%%%%%%%%%%%%%%%%%%%%%%
\section{Introduction}
%(a) riassunto sulle GL fluctuations

Within the study of superconductivity, the possibility that liquid vortex excitations survive above $T_c$, leading to local superconductivity, remains a key open problem, despite the variety of experimental and theoretical efforts.\cite{Xu2000,Cyr2009,Wang2006,Fisher1991,Iguchi,Yukalov2004}
Among all the experimental tools, the measure of the magnetization\cite{Carretta2000,Lascialfari2009,Randeira1994,Rigamonti2005} is the most straightforward, as it gives a clear diamagnetic response $M_{fl}$, as soon as the Cooper pairs are formed, provided to subtract possible paramagnetic contributions, occurring at a temperature above the onset of the fluctuations. In this regard, the Ginzburg-Landau (GL) theory predicts, for evanescent fields $H$, a diamagnetic magnetization linear in $H$ for $T\gg T_c$, and accordingly a susceptibility $\chi_{dia} \propto -\epsilon ^{D/2-2}$, where $\epsilon=(T-T_c)/T_c$ is the reduced temperature and $D$ is the system dimensionality. For a finite magnetic field and temperatures $T\rightarrow T_c(0)$, the diamagnetic part of the magnetization diverges as $(T-T_c)^{-1/2}$, and $M_{dia}\sim \sqrt{H}$. \cite{Tinkham,Larkin}\\
In addition, a relevant feature of superconducting fluctuations (SF) is the presence of an upturn field $H_{up}$ in the isothermal curves of the fluctuating magnetization $M_{fl}(H)$. In fact, while the size of fluctuating pairs $\xi(T)$ grows when the temperature approaches $T_c$, $|M_{fl}|$ shows a progressive increase. At the same time, very high magnetic fields must quench the superconducting fluctuations. The combination of the two effects leads the isothermal magnetization curves to exhibit an upturn $H_{up}$, the value of which is, for layered superconductors in the framework of the GL phenomenology, of the order of $\Phi_0/\xi^2$.\cite{Larkin,Ale2002} Therefore in optimally doped high-$T_c$ superconductors, and iron-pnictides, the upturn could be possibly detected only at very strong fields ($H \geq 10$ T), even for temperatures close enough to $T_c$.\cite{Tinkham} 

These fluctuations have been detected in several compounds, as In and Pb,\cite{Gollub} metallic nanoparticles,\cite{Ettore2006} MgB$_2$,\cite{Ale2002,Laura2005} as well as optimally doped high-$T_c$ cuprates.\cite{Ale2002_2} In fact the high-$T_c$ superconductors are ideal materials to study the SF, owing to their small coherence length $\xi$, the reduced carrier density $n_s$, the strong anisotropy $\gamma$, and high transition temperature. 

On the contrary, underdoped and overdoped cuprates show dramatic deviations from the previous behavior, namely (i) the upturn field is not ascribable to GL theory; (ii) in correspondence with the upturn in $|M_{fl}|$, the susceptibility $\chi_{dia}$ for $T\rightarrow T_c^+$ is anomalously large;\cite{Ale2002_2} and (iii) magnetic irreversibility is often found. A remarkable example of this phenomenology is given by YBa$_2$Cu$_3$O$_{6+x}$.\cite{Carretta2000}

The early interpretative attempts were due to Sewer and Beck\cite{Sewer} who described this phenomenology in terms of phase instabilities due to the formation of a vortex liquid, which can be either thermally excited, or induced by the magnetic field.
Later on, following this inspiring idea, and by taking into account terms in the free energy functional initially neglected, the model was extended,\cite{Laura2005,Ale2002_2} and the upturn in $|M_{fl}|$ at low field was justified from a phenomenological point of view. The latter model assumes the occurrence, above $T_c$, of mesoscopic islands where $|\psi|\neq 0$, but with strong phase fluctuations which inhibit the long-range coherence.\cite{Li2010} This theory successfully described the behavior of several materials,\cite{Ale2002_2, Ale2002,Giacomo2011} and it is in agreement with scanning SQUID microscopy results,\cite{Iguchi} as well as Nernst effect,\cite{Li2010,Cyr2009} and ARPES.\cite{Kondo2009}

After 2008, much attention was devoted to the iron-based superconductors. Among all the families, the 122 is one of the most promising for applications, as it is characterized by reduced thermal fluctuations, owing to a small Ginzburg number, $G_i \sim 1.5\times 10^{-5}$.\cite{Pallecchi2012} Nevertheless a few studies concerning superconducting fluctuations have been performed, one of the most noticeable being a study on a 1111 compound\cite{Giacomo2011} that showed the presence of phase fluctuations, manifested in a very similar way to cuprate superconductors. \\

In the present paper, we describe a Superconducting Quantum Interference Device (SQUID) study of the magnetization of the Ba(Fe$_{1-x}$Rh$_x$)$_2$As$_2$ (BaFeRh122) compounds. Within the 122 iron-based superconductors, this is the first study of precursor diamagnetism systematically carried out over a broad doping and magnetic field range, where the isotherm magnetization curves and their small upturn fields are discussed in the framework of the phase fluctuation model. We note that superconducting fluctuations in iron-based superconductors have been recently studied also by magnetoconductivity and magnetization, in the high field limit.\cite{Choi2009,Kim2013,Salem2009} In these cases, the critical fluctuation region has a three dimensional nature. We show that our results, at high field, agree with the literature, while the low field limit results, not yet studied on these compounds, can be explained in terms of non GL fluctuations.

%%%%%%%%%%%%%%%%%%%%%%%%%%EXPERIMENTALS%%%%%%%%%%%%%%%%%%%%%%%%%%%%%%%%%%%%%

\section{Experimental results}

\begin{figure}[htbp]
\centering
\includegraphics[width=8.2cm, keepaspectratio]{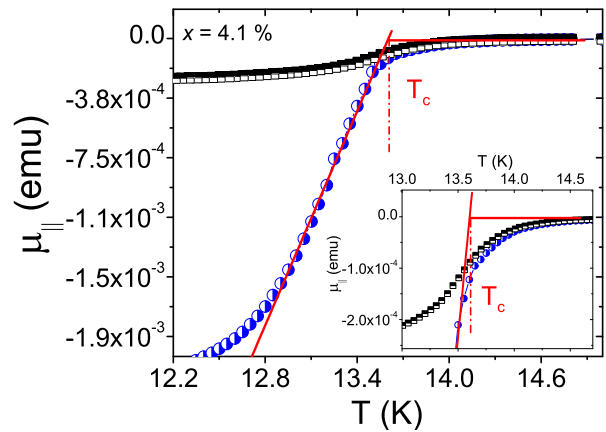}
\caption{The longitudinal magnetic moment $\mathbf{\mu}$ measured with the external field applied along the $c$ axis on the compound $x = 4.1 \%$ (UnD). The blue circles refer to the Zero-Field-Cooled data (ZFC) while the black squares refer to the Field-Cooled (FC) data. The dashed red horizontal line marks the Pauli-like trend, while the oblique line is a linear fit of the onset of superconductivity. The crossing point identifies the critical temperature $T_c$. The inset is a zoom around the critical temperature, to evidence the smoothing of the transition, which accompanies the fluctuating diamagnetism.}
\label{Fig1}
\end{figure}

Single crystals of Ba(Fe$_{1-x}$Rh$_x$)$_2$As$_2$ are grown out of self-flux, using conventional high-temperature solution growth techniques.\cite{Ni2008} 
Magnetization measurements have been performed by Superconducting Quantum Interference Device (SQUID) in the Reciprocating Sample Operation (RSO) which allows to get a better sensitivity (down to 10$^{-8}$ emu). We studied three single crystals: an underdoped (UnD) sample, with $x$ = 4.1\%
($T_c(0) = 13.6$ K), a nearly optimally doped (nOpD) sample, with $x$ = 7\% ($T_c(0) = 22.3$ K), and an overdoped (OvD) sample, with $x$ = 9.4\% ($T_c(0) = 15.1$ K).
All measurements have been done in static magnetic fields up to 7 T, parallel to the $c$ axis. The critical temperatures have been estimated from the magnetization curves versus temperature, at small fields (5 Oe), by extrapolating at $M=0$ the linear behavior of $M$ occurring below $T_c$, measured in Zero Field Cooling (ZFC), as shown in Fig. \ref{Fig1}. At a closer look, the static uniform susceptibility shows a precursor diamagnetism, at each doping (Fig. \ref{Fig1} (inset)).
Note that given the sharpness of the curves in Fig. \ref{Fig1}, we exclude the possibility of an asymmetric distribution of $T_c$.\cite{Mosqueira2011_2}

%%%%%%%%%%%%%%%%%%%%%%%%%%%%%%%
%%%%%%%%%%%%%%%%%%%%%%%%%%%%%%%%%

\begin{figure}[htbp]
\centering
\includegraphics[width=11.2cm, keepaspectratio]{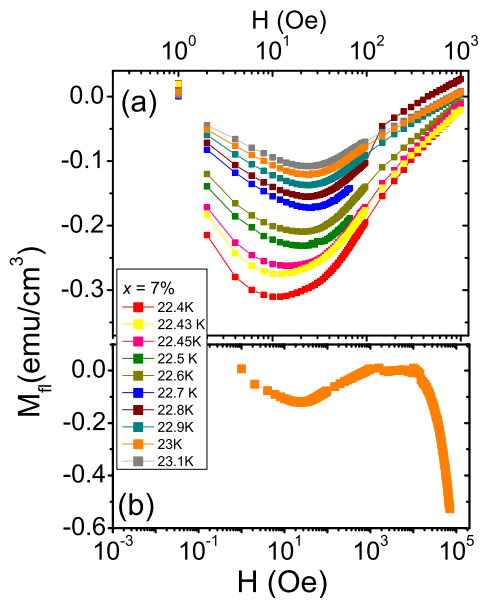}
\caption{\textbf{(a)} The fluctuating magnetization as obtained by equation (\ref{fl}), for $x=7$ \% (nOpD) compound, with $T_c(0)=22.3$ K. The upturn, as well as its trend with temperature are noticeable. \textbf{(b)} $M_{fl}$ Vs H at 23 K for the same sample: a strong diamagnetic contribution arises at larger fields, reminiscent of a second upturn. The first upturn has been added for comparison. 
}
\label{Fig2}
\end{figure}

After obtaining $T_c(0)$, the magnetization has been measured as a function of the field (isotherm) at different temperatures: high resolution scans have been performed for several isotherms, above $T_c$, generally by 0.03 K increments, in steps of 2 Oe, up to 1000 Oe. Additional scans at a temperature well above $T_c$ allow to determine the paramagnetic signal of the normal state, and the background contributions from the sample holder. In fact the measured magnetization $M$ is a sum of the fluctuation magnetization $M_{fl}$, the normal state $M_{n}$, and the sample holder $M_{sh}$ contributions. Since the latest are nearly temperature-independent, in the temperature range observed, the fluctuation magnetization is estimated by the subtraction:
\begin{equation}
M_{fl}(T,H)=M(T,H)-(M_n+M_{sh}).\label{fl}
\end{equation}
Some isothermal magnetization curves are also measured below $T_c$ and well above $T_c$, for comparison. 
The isothermal curves $M_{fl}(H)$ of all the measured specimens show a clear upturn, visible for $x$=7\% sample in Fig. \ref{Fig2} \textbf{(a)}, in the range of 10 - 50 Oe, while the application of fields in the Tesla range allows to single out the onset of a second strong diamagnetic response (Fig. \ref{Fig2} \textbf{(b)}), which will be discussed subsequently.
%%%%%%%%%%%%%%%%%%%%%%%%%%%%%%%
%%%%%%%%%%%%%%%%%%%%%%%%%%%%%%%%%

\section{Discussion and interpretation of the results}

\begin{figure}[htbp]
\centering
\includegraphics[width=10cm, keepaspectratio]{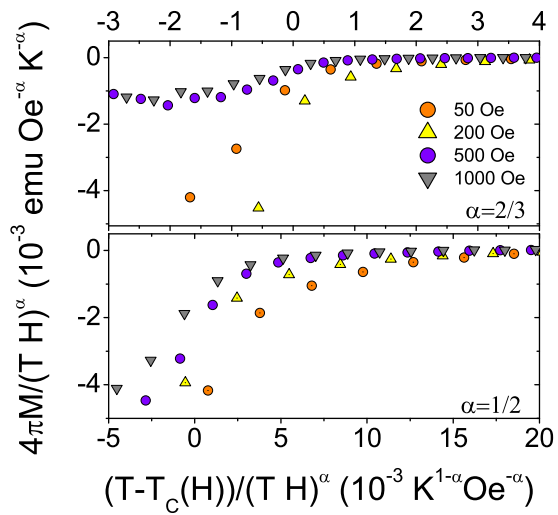}
\caption{Reduced magnetization as a function of the reduced temperature, for $x=7\%$ (nOpD) sample. The top panel refers to the 3D scaling law, i.e. $\alpha=2/3$, while the bottom panel refers to the 2D scaling, i.e. $\alpha=1/2$. In both cases the scaling law does not apply, below $\leq$ 1 kOe.}
\label{Fig4a}
\end{figure}

The origin of the low field upturn occurring in $M_{fl}(H)$ (Fig. \ref{Fig2}) cannot be related to Gaussian GL fluctuations. In fact, for a coherence length $\xi \sim 20 -30 $ \AA\ (see Ref. \onlinecite{Ni2009} for comparison) and $\epsilon \sim 0.02$, the upturn is expected in the range of 10 T, whereas our upturn field is four orders of magnitude smaller.
Moreover, the 3D and 2D scaling laws, consistent with the GL theory,\cite{Kamal94,Welp91} fail within the small field limit, \textit{i.e. below 1 kOe}, as it is shown in Fig. \ref{Fig4a}. Indeed according to the scaling, universal curves of the form $M/(TH)^{\alpha}$ vs $[T- T_c(H)]/(TH)^{\alpha}$ are expected, where $\alpha=3/2$ ($1/2$) for a 3D (2D) system. These evidences support the idea that a different fluctuating mechanism, not ascribable to the Aslamazov-Larkin fluctuations, has to be found. We note that our data are not in disagreement with recent results on iron-pnictides, since we are dealing with a low field region. On the contrary at larger fields, the 3D scaling will be recovered, as shown later on in Fig \ref{Fig7}. Nevertheless we believe that the former field range, i. e. low fields, presents more interesting results, that we will discuss in the following.

%%%%%%%%%%%%%%%%%%%%%%%%%%%%%%%
%%%%%%%%%%%%%%%%%%%%%%%%%%%%%%%%%

\begin{figure}[htbp]
\centering
\includegraphics[width=7cm, keepaspectratio]{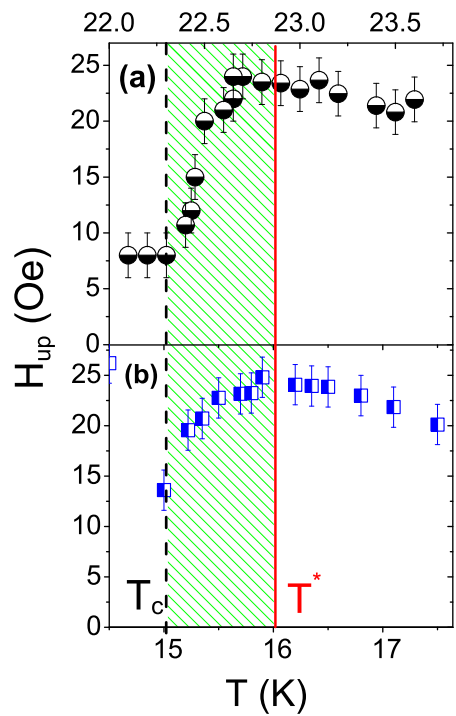}
\caption{\textbf{(a)} The upturn field as a function of temperature for $x = 7 \%$ (nOpD), derived by calculating the derivative of $M_{fl}$ with respect to the field. \textbf{(b)} The upturn field for $x = 9.4 \%$ (OvD) sample, where the fluctuation region is wider. The phase fluctuation region is marked by a green area, and the "diffuse- $T_c$" region is found at $T>T^*$, where $T^*$ is marked by the red line (see text).}
\label{Fig4}
\end{figure}

One might ascribe the extra diamagnetism to sample inhomogeneity, that would give rise to a diffuse transition with $T_c^{loc}(\mathbf{r})>T_c^{bulk}$.\cite{Laura2003,Cabo2006} If this was the case, the upturn field would be significantly decreased with the increase of the temperature, in a $H_{c1}$-like manner. On the contrary, here the upturn field increases with the increase of temperature, although in a narrow temperature range above $T_c$ (Fig. \ref{Fig4}), while at higher temperatures (T$>T_c$+0.6 K for $x$=7\% sample, and T$>T_c$+ 1 K for $x$=9.4\% sample) H$_{up}$ tends to slowly decrease. In the region where a smooth decrease of $H_{up}$ is observed, we do not exclude the presence of a slight diffuse transition together with preformed superconducting droplets, but we believe that the former is not the predominant physics effect. We note that, due to the sharpness of the transition, a non-asymmetric distribution of $T_c$ recently claimed,\cite{Mosqueira2011,Mosqueira2011_2} can be excluded as the cause of $H_{up}$ increase with temperature.\\

%%%%%%%%%%%%%%%%%%%%%%%%%%%%%%%
%%%%%%%%%%%%%%%%%%%%%%%%%%%%%%%%%

To provide a quantitative interpretation of our data, we assumed that in a phase fluctuation scenario, the amplitude of the order parameter $|\Psi|$ is frozen at a non-zero value, while the phase $\theta$ fluctuates in time, preventing the formation of long-range superconductivity. Such model predicts the presence, above the bulk $T_c$, of superconducting "droplets", characterized by the formation of metastable, i.e. fluctuating in phase, thermally activated vortex loops $n_{th}$ and field induced $n_{H}$ vortex lines, whose nature depends on the field and temperature range chosen above $T_c$.

\begin{figure}[htbp]
\centering
\includegraphics[width=10.4cm, keepaspectratio]{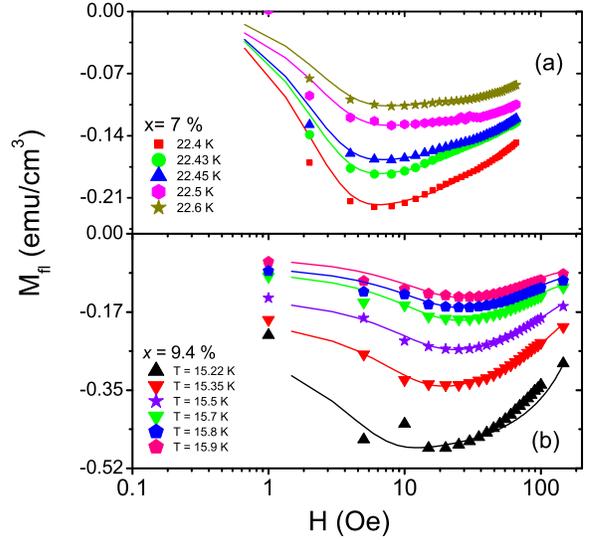}
\caption{The fluctuating magnetization, as a function of the field, for sample $x=7$ \% (nOpD) \textbf{(a)}, and $x=9.4$ \% (OvD) \textbf{(b)}. The solid lines are the fit to Eq.(4).}
\label{Laura}
\end{figure}

\begin{figure}[htbp]
\centering
\includegraphics[height=6cm]{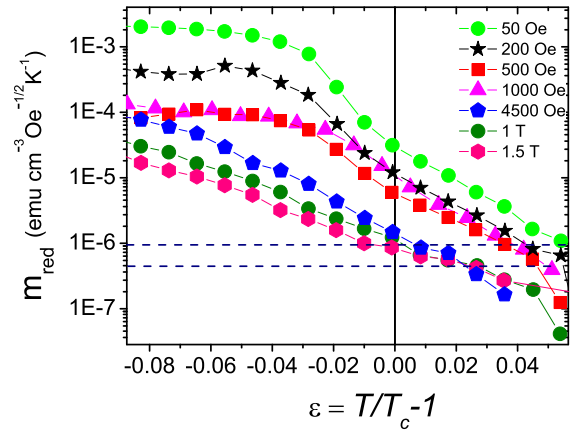}
\caption{The reduced magnetization as a function of the reduced temperature $\epsilon=(T-T_c)/T_c$, for fields below 2 T, for $x = 7\%$ (nOpD).\cite{nota2} At small fields, the curves show a parallel behavior around $\epsilon=0$. At larger fields ($H>500$ Oe), the data cross at values included into the two dashed lines. Such values are in agreement with the GL theory for 3D \textit{XY} systems,\cite{Junod1998} when an anisotropy ratio of the value around 1-2 is assumed.} 
\label{Fig7}
\end{figure}

For small fields, and $T_c<T<T_c$+0.6 K for $x$=7\% sample, $T_c<T<T_c$+1 K for $x$=9.4\% sample, a diamagnetic effect due to metastable vortices can be evinced. Within this scenario, the temperature dependence of the susceptibility is controlled by the vortex loops density\cite{Ale2003} 
\begin{equation}
n_{th}(T,H)=n_0e^{-E_0(1+2n)/k_B T(1+\delta (h)^2)},\label{act}
\end{equation}
where the activation energy $E_0$ depends on the number of layers $n$ involved in the pancakes and $\delta=\pi^2(J_{\parallel}/k_BT)$, where $k_B$ is the Boltzmann constant and $J_{\parallel}$ is the Josephson coupling along the planes. In the last equation, $h=H/H^*$ where $H^*=\Phi_0/\mathcal{L}^2$ is a magnetic field which takes into account the dimension of the superconducting droplets, and $\Phi_0$ is the magnetic quantum flux.\\
In the opposite limit of high fields, but still below 1 kOe, a different diamagnetic contribution can be singled out: here the field induced vortices are dominant, and the $n_H$ density is given by
\begin{equation}
n_H=\frac{H}{\Phi_0}=\frac{H}{H^*\mathcal{L}^2}.
\end{equation}
Taking into account both low and high fields limits for T$>T_c$, the second derivative of the free energy with respect to $H$ yields the field-dependent susceptibility:\cite{Ale2002_2}
\begin{eqnarray}
\chi(T,H)&= &-\frac{k_B T }{s\Phi_0^2}\frac{1}{1+2n}\frac{(1+\delta(h)^2)^2}{n_v}-\notag \\
& & \frac{s^2 \gamma^2(1+n)}{1+2n}[1+\delta(h)^2]+\notag \\
& &\frac{47 \mathcal{L}^2}{540}\frac{J_{\parallel}}{s}\left( \frac{2\pi}{\Phi_0}\right)^2\delta(h)^2 
\end{eqnarray}
where $s$ is the interlayer distance, $\gamma=\xi_{ab}/\xi_c$ is the anisotropy ratio, and $n_v=n_{th}+n_H$.
After fitting the data to the numerical integration of Eq. (4), we found two contributions to the diamagnetism, coming respectively from phase fluctuations of the order parameter and from the effects of a diffuse transition, whose existence is due to different parts of the compounds experiencing different superconducting transition temperatures. Firstly we note that the fit to the raw data gave satisfactory results, with a slight mismatch in the field range $H>H_{up}$, as already observed in Ref. \onlinecite{Giacomo2011}. The fit quality could be further refined by subtracting a curve chosen at $T^*$, defined as the temperature where the upturn field begins to decrease (see Fig. \ref{Fig4}). We remark that this rudimentary approach is just a first attempt to take into account the simultaneous presence of phase fluctuations and diffuse transition. Finally, the fit to Eq.(4) turns out reasonably good (Fig. \ref{Laura}). 
%%%%%%%%%%%%%%%%%%%%%%%%%%%%%%%%%%%%%%%%%%%%%%%%%%%%%%%%%%%%%%%%%%
%%%%%%%%%%%%%%%%%%%%%%%%%%%%%%%%%%%%%%%%%%%%%
%Even though more accurate analysis is needed, this evidence suggests a simple approach to the problem of treating the coexistence of SF with a distribution of transition temperatures. 

In the fitting procedure, the interlayer distance has been set to s=6 \AA\ according to the structural data,\cite{Ni2008} while the anisotropy ratio has been taken in the range $\gamma=2-3$. This small anisotropy implies a large number of layers involved in the pancakes. Indeed, the best fit to the isothermal magnetization curves, at $T$ = 22.4 K for $x$=7\% and 15.35 K for $x$=9.4\%, gives for the expression $E_0(1+2n)/k_BT$ the value 96 in the former case, and 76 in the latter, corresponding respectively to $n\sim 11$ and $n\sim 9$.\\
Moreover the activation energy in Eq.(\ref{act}) turns out $E_0\sim 2k_BT_c$ in the two cases, which is of the same order of magnitude usually estimated in YBCO ($E_0 \sim 10k_BT_c$).\cite{Ale2002} By comparison, YBCO and Sm-based cuprates\cite{Ettore2010} showed $n$ = 3, as expected from their high anisotropy. \\
In addition, the characteristic field $H^*$ gives an estimate of the average size $\mathcal{L}$ of the superconducting regions. In Table \ref{Tb1} the fitting parameters, for the nearly optimally doped and overdoped compound, are reported. Also the underdoped sample displays a similar phenomenology. However, because of the small sample size, i. e. small magnetic moment, a reliable fit could be hardly achieved.\\
From the fit, the pre-exponential factor $n_0$ turns out to be 2.5$\times 10^{19}$ cm$^{-3}$ for $x=7$ \%, and 1.5$\times 10^{20}$ cm$^{-3}$ for $x=9.4$ \%. \\
%-------------------- Table 1 ------------------------------

\begin{table}[ht]
\caption{Fit results for samples $x=$7\% and 9.4\%. $\mathcal{N}$ represents the volumetric density of mesoscopic islands, of surface $\mathcal{L}^2$.} %
\vspace{0.3cm}
\centering % used for centering table
\begin{tabular}{c c c } % centered columns (4 columns)
\hline
\hline
$\epsilon$ & $\mathcal{N}\times 10^{14}$(cm$^{-3})$ & $\mathcal{L}$ (nm)\\ [0.5ex] % inserts table
\hline
\hline
	& $\mathbf{x=7 \%}$ &    \\
\hline	

0.0048	& 3.0234 & 580   \\
\hline
0.0058	& 2.5426 &	574  \\
\hline
0.0067	& 2.5176 	& 550	 \\
\hline
0.0089 & 2.0198	& 530 \\
\hline
0.01345	& 1.8153 & 510	\\
\hline % [1ex] adds vertical space
\hline %inserts single line

	& $\mathbf{x=9.4 \%}$ &    \\
\hline

0.01466	& 9.0625 &	480 \\
\hline
0.023	& 1.2706	&310	\\
\hline
0.033	& 1.1784	&270\\
\hline
0.0466  & 1.10283 &	260	\\
\hline
0.0533 &	0.8610 &	260	\\
\hline
0.06	&0.7077	&260	\\
\hline % [1ex] adds vertical space
\hline %inserts single line
\end{tabular}
\label{Tb1} % is used to refer this table in the text
\end{table}
%--------------------- end of the table ------------------------

In addition to the above results, at larger fields, the magnetization reveals a second strong diamagnetic response (Fig. \ref{Fig2}\textbf{(b)}), whose temperature dependence agrees with the scaling laws derived from the GL theory, in presence of a small anisotropy. Indeed, the reduced magnetization curves $m_{red}=\frac{M}{\sqrt{H} T_c}\label{mred}$ cross at a universal value (Fig. \ref{Fig7}):
\begin{equation}
m_{red}(T_c)=\frac{M(T_c)}{\sqrt{H} T_c}=\frac{k_B}{\Phi_0^{3/2}}m_3(\infty)\gamma \label{scaling},
\end{equation}
where $k_B$ is the Boltzmann constant, and $m_3(\infty)=-0.323$.\cite{Hubbard,Junod1998} Interestingly, the crossing point implies an anisotropy of $\gamma=1-2$ (Fig. \ref{Fig7}), in fairly good agreement with the literature.\cite{Ni2008,Ni2009} \\

The existence of anisotropic GL or Aslamazov-Larkin SF, in the 122 family of iron-based superconductors, has been observed also by means of other experimental techniques. Mosqueira et \textit{al.} recently found GL fluctuations in Ba$_{1-x}$K$_x$Fe$_2$As$_2$,\cite{Mosqueira2011} in agreement with a 3D anisotropic scenario in the high field range. In addition, a Raman study on a sample of Ca$_4$Al$_2$O$_{5.7}$Fe$_2$As$_2$ showed an anomaly at 60 K $\sim$ 2 $T_c$, which has been ascribed to a strong coupling of the observed phonon mode with the superconducting order parameter fluctuations.\cite{Kumar2012}\\

Before concluding, we note that a magnetic irreversibility in the isothermal magnetization was observed, above $T_c$ (data not shown). This effect suggests the entrance of magnetic flux, in correspondence with precursor diamagnetic islands. It is not clear at the moment whether the irreversibility
is due to vortices entering the region of the sample characterized by a local distribution of $T_c$, or to field-induced vortices entering the metastable islands discussed in the text; hence we do not discuss it further. \\

%%%%%%%%%%%%%%%%%%%%%%%%%%%%%%%%%%%%%%%%%%%%%%%%%%%%
%%%%%%%%%%%%%%%%%%%%%%%%%%%%%%%%%%%%%%

\section{Summarizing Remarks and Conclusions}

The diamagnetic response of BaFeRh122 crystals has been studied by magnetization measurements, over a broad range of doping content. At low magnetic fields, and $T_c<T<T_c+\delta$ ($\delta$=0.6 K for $x$=7\%, and $\delta$=1 K for $x$=9.4\%) the experimental findings cannot be described within the Ginzburg-Landau theory, while they are consistent with the "phase fluctuating scenario". In this framework, unconventional SF are supposed to develop into precursor superconducting islands, where the amplitude of the order parameter is frozen, while the long-range phase coherence associated with the bulk superconducting state is prevented by strong fluctuations of the phase. At higher temperatures, i.e. T$>T_c+\delta$, the effect of diffuse transition is likely responsible for the decrease of the upturn field. \\
On the other hand, when strong fields are applied, the Aslamazov-Larkin fluctuation scenario is restored and the experimental findings agree with the 3D \textit{XY} universality class. In fact the crossing of the reduced magnetization at $T_c$, supports our thesis. Such dichotomy seems peculiar of the iron-based superconductors, since analogous results were found in a member of the 1111 family.\cite{Giacomo2011} 

Finally, it can be remarked that this phenomenology occurs in a narrow fluctuation region, which makes more difficult to study the effects accompanying the SF, at variance with the more noticeably phenomena observed in the cuprates. Still, the experimental evidences displayed in the present paper clearly support the presence of superconducting phase fluctuations, in the iron-based superconductors.

\section*{Acknowledgments}

P. Carretta and A. Rigamonti are gratefully acknowledged for useful discussions.
The research activity in Pavia was supported by Fondazione Cariplo
(Research Grant No. 2011-0266). Work done in Ames Laboratory (P.C.C.) was supported by the U.S. DOE, Basic Energy Sciences Office, Division of Materials Sciences and Engineering Division under contract No. DE-AC02-07CH11358. We acknowledge N. Ni, S. Ran and A. Thaler for assistance in sample preparation.

%%%%%%%%%%%%%%%%%%%%%%%%%%%%%


\begin{references}


\bibitem{Xu2000} Z. A. Xu, N. P. Ong, Y. Wang, T. Kakeshita, and S. Uchida, Nature
 (London) \textbf{406}, 486 (2000).
\bibitem{Cyr2009} Olivier Cyr-Choini\'ere, R. Daou, Francis Lalibert\'e, David LeBoeuf, Nicolas Doiron-Leyraud, J. Chang, J.-Q. Yan, J.-G. Cheng, J.-S. Zhou, J. B. Goodenough, S. Pyon, T. Takayama, H. Takagi, Y. Tanaka and Louis Taillefer, Nature \textbf{458}, 743 (2009).
\bibitem{Wang2006} Y. Wang, L. Li and N. P. Ong, Phys. Rev. B \textbf{73}, 024510 (2006). 

\bibitem{Iguchi} I. Iguchi, T. Yamaguchi, and A. Sugimoto, Nature (London) \textbf{412},
 420 (2001).
\bibitem{Yukalov2004} V. I. Yukalov and E. P. Yukalova, Phys. Rev. B, \textbf{70}, 224516 (2004). V. I. Yukalov, Int. J. Mod. Phys. B \textbf{6}, 91 (1992).
\bibitem{Fisher1991} D. S. Fisher, M. P. A. Fisher and D. A. Huse, Phys. Rev. B \textbf{43}, 130 (1991).
\bibitem{Carretta2000} P. Carretta, A. Lascialfari, A. Rigamonti, A. Rosso, and A. Varlamov, Phys. Rev. B \textbf{61}, 12420 (2000).
\bibitem{Lascialfari2009} A. Lascialfari, A. Rigamonti, E. Bernardi, M. Corti, A. Gauzzi and J. C. Villegier, Phys. Rev. B \textbf{80}, 104505 (2009).
\bibitem{Rigamonti2005} A. Rigamonti, A. Lascialfari, L. Roman\'o\, A. Varlamov, and I. Zucca, Journal of Superconductivity \textbf{896}, 1107 (2005).
\bibitem{Randeira1994} M. Randeira and A. A. Varlamov, Phys. Rev. B \textbf{50}, 10401(R) (1994).
\bibitem{Tinkham} M. Tinkham, \textit{Introduction to Superconductivity}, Second Edition,  Dover (1996).%22
\bibitem{Larkin} A. I. Larkin and A. A. Varlamov, \textit{Theory of Fluctuations
in Superconductors}, Clarendon Press, Oxford (2005).
\bibitem{Ale2002} A. Lascialfari, T. Mishonov, A. Rigamonti, P. Tedesco, and A. Varlamov, Phys. Rev. B \textbf{65}, 180501(R) (2002).
\bibitem{Gollub} J. P. Gollub, M. R. Beasley, R. Callarotti and M. Tinkham, Phys. Rev. B \textbf{7}, 3039 (1973).
\bibitem{Ettore2006} E. Bernardi, A. Lascialfari, A. Rigamonti, L. Roman\'o, V. Iannotti, G. Ausanio and C. Luponio, Phys. Rev. B \textbf{74}, 134509 (2006).
\bibitem{Laura2005} L. Roman\'o, A. Lascialfari, A. Rigamonti and I. Zucca, Phys. Rev. Lett. \textbf{94}, 247001 (2005).
\bibitem{Ale2002_2} A. Lascialfari, A. Rigamonti, L. Roman\'o, P. Tedesco and D. Embriaco, Phys. Rev. B \textbf{65}, 144523 (2002).
\bibitem{Sewer} A. Sewer and H. Beck, Phys. Rev. B \textbf{64}, 014510 (2001).
\bibitem{Li2010} L. Li, Y. Wang, S. Komiya, S. Ono, Y. Ando, G. D. Gu, and N. P . Ong, Phys. Rev. B \textbf{81}, 054510(R) (2010).
\bibitem{Giacomo2011} G. Prando, A. Lascialfari, A. Rigamonti, L. Roman\'o, S. Sanna, M. Putti, and M. Tropeano, Phys. Rev. B \textbf{84}, 064507 (2011).
\bibitem{Kondo2009} T. Kondo, R. Khasanov, T. Takeuchi, J. Schmalian and A. Kaminski, Nature (London) \textbf{457}, 297 (2009).

\bibitem{Pallecchi2012} I. Pallecchi and M. Tropeano and G. Lamura and M. Pani and M. Palombo and A. Palenzona and M. Putti, Phys. C \textbf{482}, 68 (2012).
\bibitem{Choi2009} C. Choi, S. Hyun Kim, K.-Y. Choi, M.-H. Jung, S.-I. Lee, X. F. Wang, X. H. Chen and
X. L. Wang, Supercond. Sci. Technol. \textbf{22}, 105016 (2009).
\bibitem{Kim2013} S. Hyum Kim \textit{et al.}, J. Appl. Phys. \textbf{108}, 063916 (2010).
\bibitem{Salem2009} S. Salem-Sugui \textit{et al.}, Phys. Rev. B \textbf{80}, 014518 (2009).

\bibitem{Ni2008} N. Ni, M. E. Tillman, J.-Q. Yan, A. Kracher, S. T. Hannahs,
S. L. Bud'ko, and P. C. Canfield, Phys. Rev. B \textbf{78}, 214515 (2008).
\bibitem{Mosqueira2011_2} J. Mosqueira, J. D. Dancausa, and F. Vidal, Phys. Rev. B \textbf{84}, 174518 (2011).
\bibitem{Ni2009} N. Ni, A. Thaler, A. Kracher, J. Q. Yan, S. L. Bud'ko, and P. C. Canfield, Phys. Rev. B \textbf{80}, 024511 (2009).
\bibitem{Kamal94} S. Kamal, D. A. Bonn, N. Goldenfeld, P. J. Hirschfeld, R. Liang, and W. N. Hardy, Phys. Rev. Lett. \textbf{73}, 1845 (1994).
\bibitem{Welp91} U. Welp, S. Fleshier, W. K. Kwok, R. A. Klemm, V. M. Vinokur, J. Downey, B. Veal,
and G. W. Crabtree, Phys. Rev. Lett. \textbf{67}, 2180 (1991).
\bibitem{Laura2003} L. Roman\'o, Int. J. Mod. Phys. B \textbf{17}, 423 (2003).
\bibitem{Cabo2006} L. Cabo, F. Soto, M. Ruibal, J. Mosqueira, and F. Vidal, Phys. Rev. B \textbf{73}, 184520 (2006).
\bibitem{Mosqueira2011} J. Mosqueira, J. D. Dancausa, F. Vidal, S. Salem-Sugui Jr., A. D.
Alvarenga, H.-Q. Luo, Z.-S. Wang, and H.-H. Wen, Phys. Rev. B
\textbf{83}, 094519 (2011).
\bibitem{Ale2003} A. Lascialfari, A. Rigamonti, L. Roman\'o, A. A. Varlamov and I. Zucca, Phys. Rev. B \textbf{68}, 100505(R) (2003).
\bibitem{Ettore2010} E. Bernardi,  A. Lascialfari, A. Rigamonti, L. Roman\'o, M. Scavini and C. Oliva, Phys. Rev. B \textbf{81}, 064502 (2010).
\bibitem{Junod1998} A. Junod, J.-Y. Genoud, G. Triscone and T. Schneider, Physica C \textbf{294}, 115 (1998).
\bibitem{Hubbard} M. A. Hubbard, M. B. Salamon and B. W. Veal, Physica C \textbf{259}, 309 (1996).

\bibitem{Kumar2012} P. Kumar, A. Bera, D. V. S. Muthu, P. M. Shirage, A. Iyo, A. K. Sood, Appl. Phys. Lett. \textbf{100}, 222602 (2012).
\bibitem{nota2} These measurements were carried out in Field Cooled (FC), and the diamagnetic magnetization was estimated by subtracting from the raw data the high temperature (i.e. normal phase) contribution, the latest being evaluated by a polynomial fit.







\end{references}
\end{document}